*Quadrimechanica*
*Entropy: the basis of mechanics.*

*Roberto Assumpção*
*PUC-Minas, Av. Pe.Francis C. Cox, Poços de Caldas MG  37701-355, Brasil*
*assumpcao@pucpcaldas.br*## # - Abstract

Classical, Quantum, Relativistic and Statistical: the four branches of mechanics. However, the *Quattro Donna* of Physics disagree even about the entities that are supposed to be fundamental, such as space, matter and time. In order to search for an union, this contribution considers Indeterminacy in a Classical context and takes a path based on a division that could be named *factual* as opposed to the conventional approach of *cause and effects*. This sets back a primary division between "observable entities" and "measurable quantities". Recovering the primordial Classical experiment and describing the "Free Fall" in terms of distinct times – the relaxation time of the ("falling") 'light' bodies ( $m_1$ , $m_2$,...); the characteristic time of the 'heavy' (M) body and the rate of change of the mechanical system [M, $m_1$, $m_2$, ...]– the procedure led to conclude that Newtonian Mechanics extracted data from a *gedanken* experiment. Developing the concept of evolution of states (Entropy), an argument centred on *Observability* favours a uniform formulation, at the cost of non-linear "Classical" equations.## # 1 – Introduction

Historically, a label could be set in the year of 1642 – birth of Newton and passage of Galileo – considering this date as a mark of the transition from a philosophical qualitative description of observed facts to a physical quantitative representation of experimental results; in short, it is the birth of controlled laboratory methods. The primordial experiment of Mechanics [1] follows the procedure of measurement and repetition and is pictured in Figure 1: a body $m_1$ is abandoned near the proximity of a body M and the "rate of fall" determined somehow; however, to gain confidence on the result, it is necessary to repeat the experiment, preferably employing a distinct body, say $m_2$ .

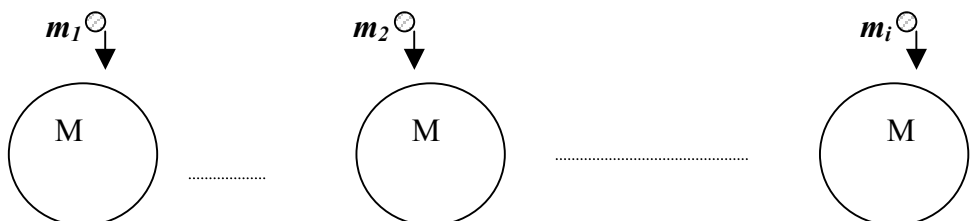

**Figure 1 – The Classical Experiment**

The result indicates a proportionality of the ratio of masses to the ratio of accelerations,

$$\frac{m_1}{M} \propto \frac{\dot{V}}{\dot{v}_1} \quad\ldots\ldots\quad \frac{m_2}{M} \propto \frac{\dot{V}}{\dot{v}_2} \quad\ldots\ldots\quad \frac{m_i}{M} \propto \frac{\dot{V}}{\dot{v}_i}$$

where the pair ($m_i, v_i$) refers to the "light" body whereas ($M, V$) to the heavier or more massive body.

The general representation of the classical experiment is:

$$\frac{m_i}{m_j} = h \times \frac{\dot{v}_j}{\dot{v}_i} \quad (1)$$

where $h$, the 'constant of proportionality', assumes the value $h = -1$ and, according to Classical mechanics, leads to the second ($P_i = m_i v_i$) and to the third laws ($F_i = -F_j$).

However, figure 1 seems unrealistic once how could one generate, indefinitely, masses $m_i$ to experiment (balance) with M ? A distinct (view) picture of the Classical experiment is shown bellow:

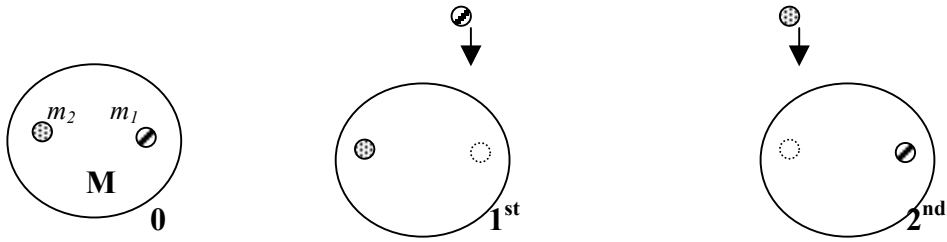

**Figure – 2 :** Initial, first and second stages of the mechanical experiment.

Comparing figures 1 and 2 we note that there is no place for $m_2$ (other than on the surface of M) when $m_1$ is "falling". Thus figure 1 represents a *gedanken* experiment, in a pure quantum-mechanical sense that it could not be carried out. In other words, the transition from the philosophical qualitative description of facts to a quantitative representation of experimental results was inconsistently conducted by Classical mechanics, leaving *the problem of observation* to Quantum and Relativistic mechanics.

Now figure 2 pictures a 3-body problem which, as far as experimental methods are concerned, is the minimum necessary arrangement to assure consistent results; according to experimental data, each stage gives:

$$1^{st}) - \frac{m_1}{M+m_2} \approx \frac{\dot{V}_2}{\dot{v}_1} \quad ; \quad 2^{nd}) - \frac{m_2}{M+m_1} \approx \frac{\dot{V}_1}{\dot{v}_2}$$

where $\dot{V}_i$ is the acceleration of ($M + m_i$) and $\dot{v}_i$ the acceleration of $m_i$; the general representation of the experiment becomes:

$$\frac{m_1}{m_2} \approx \{ \frac{M+m_2}{M+m_1} \times \frac{\dot{V}_2}{\dot{V}_1} \} \times \frac{\dot{v}_2}{\dot{v}_1} \quad (2)$$

According to this relation, the 'constant of proportionality' $h$ between the ratio of masses and the ratio of accelerations assumes the value:

$$h \equiv \{ \frac{M+m_2}{M+m_1} \times \frac{\dot{V}_2}{\dot{V}_1} \} \quad (3)$$

This way, the "free fall" becomes a 3–body experiment and $h$ reveals a dependency on the third part. A number of interpretations result: the current one, indicating no dependency on M, once $m_i$ appears "falling" just in the presence of $m_j$ – equation (1), meaning no experiment at all ! Independence on $m_j$ indicates no repetition, or a result based on a single data; another possibility is that the ( full ) value of *h* can be obtained from the analysis of the experimental data.

# # 2 – *Analysis of the "Classical" experiment*

**A)-** Let $m_2 \approx m_1 + \delta M$ ; elimination of **$m_2$** from equation ( 3 ) gives:

$$h \approx \frac{M + m_1 + \delta M}{M + m_1} \times \frac{\dot{V}_2}{\dot{V}_1} \quad \text{or} \quad h \approx \left(1 + \frac{\delta M}{M + m_1}\right) \times \frac{\dot{V}_2}{\dot{V}_1}$$

Now, taking $M \gg m_1$,

$$h \approx \left(1 + \frac{\delta M}{M}\right) \times \frac{\dot{V}_2}{\dot{V}_1} \quad ; \text{defining the ratio of masses } \alpha \equiv \frac{\delta M}{M}, \quad h \text{ is given by :}$$

$$h(\alpha) \approx (1 + \alpha) \times \frac{\dot{V}_2}{\dot{V}_1} \quad \textbf{( A.1)}$$

Conversely, elimination of **$m_1$** gives:

$$m_1 \approx m_2 - \delta M$$

so that,

$$h \approx \frac{M + m_2}{M + m_2 - \delta M} \times \frac{\dot{V}_2}{\dot{V}_1} \quad \text{that is:}$$

$$h(\alpha) \approx \frac{1}{(1 - \alpha)} \times \frac{\dot{V}_2}{\dot{V}_1} \quad \textbf{( A.2 )}$$

Results ( A.1) and (A.2) appear contradictory; an analogous situation happens when body 1 is heavier than body 2:

**B)-** $m_1 \approx m_2 + \delta M$

Elimination of **$m_1$** gives**:**

$$h \approx \left(\frac{M + m_2}{M + m_2 + \delta M}\right) \times \frac{\dot{V}_2}{\dot{V}_1} \text{, that is } h \approx \left(\frac{1}{1 + \delta M/M}\right) \frac{\dot{V}_2}{\dot{V}_1} \text{ or, according to the definition of } \alpha,$$

$$h(\alpha) \approx \left(\frac{1}{1 + \alpha}\right) \frac{\dot{V}_2}{\dot{V}_1} \quad \textbf{( B.1 )}$$

Conversely, elimination of **$m_2$** gives: $m_2 \approx m_1 - \delta M$ **;** substituting in **( 3 )** leads to,

$$h \approx \left(\frac{M + m_1 - \delta M}{M + m_1}\right) \times \frac{\dot{V}_2}{\dot{V}_1} \quad \text{or} \quad h \approx (1 - \delta M/M) \frac{\dot{V}_2}{\dot{V}_1} \text{, that is,}$$

$$h(\alpha) \approx (1 - \alpha) \frac{\dot{V}_2}{\dot{V}_1} \quad \textbf{( B.2 )}$$

Relations A.1 to B.2 represent four distinct results for the "same" experiment; however, these are not precisely the same, once addition of a mass (δM) to *$m_1$* is physically distinct from subtracting a mass (δM) from *$m_2$*.

Classical mechanics may argument that $h(\alpha)$ is constant, that all values converge to $-1$; moreover, a direct experiment relating bodies $(M + m_2)$ to $(M + m_1)$ is also virtual. However, the distinction of experiments is guaranteed by $\alpha$, not by employing different lesser bodies $m_1$ and $m_2$ in successive stages.

The argument of Quantum mechanics could be that the particular value $h(\alpha)$ assumes is irrelevant, since this function is not an observable; so the path followed by the light bodies $m_i$ from the excited stages to the ground state can not be described by the theory. Relativistic mechanics may admit such a function once a connection between masses and velocities do exist, but will argue that the effect can only be observed at relativistic velocities. Statistical mechanics is the only that may consider the ambiguity and treat $h(\alpha)$ as a description of the path followed by the masses, a function that is sensitive to the order of experimentation and takes the responsibility for the fact that ideal isolation of the two bodies $m_1$ and $m_2$ could not be achieved.

This last argument will now be followed, in a study of the evolution of the mechanical states of the system $[M \mid m_1 \mid m_2]$.

...... *** ......

# 3 – *Evolution of States:*

Figure 3 pictures the evolution of the masses M, $m_1$ and $m_2$. Three times can be distinguished: the relaxation time of the bodies $m_1$ and $m_2$ or the elapsed time of "fall" ($\delta t_i$) ; the time interval between observations, or the rate of change of the system ($\delta t$). The third time is the *proper time* of M, $\Delta t$. $\Delta t$ is also the proper time of $m_1$ and $m_2$ when these are simultaneous with M, such as in stage $S_o$.

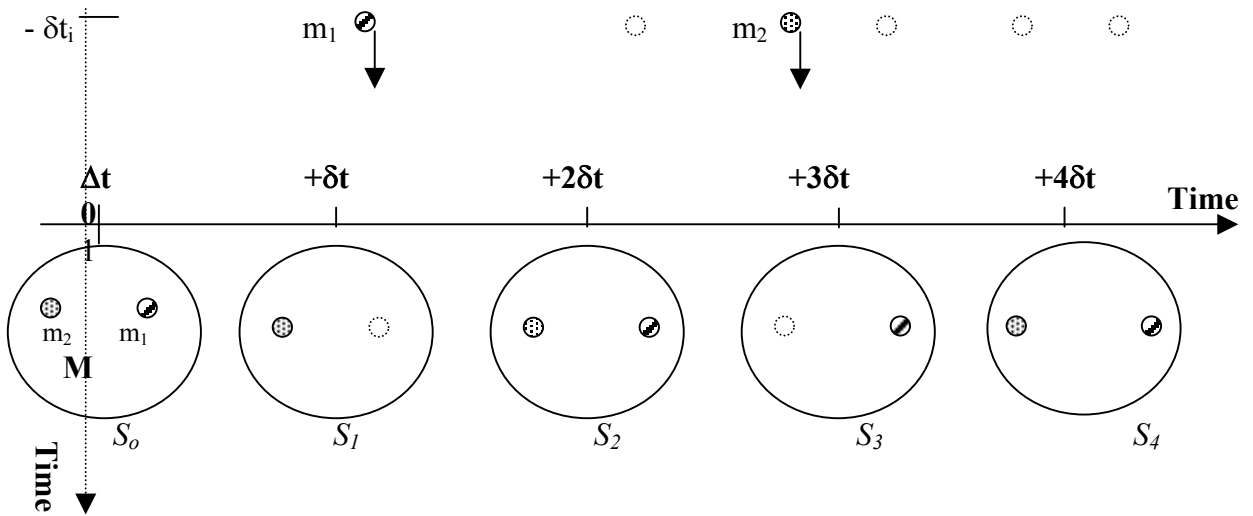

**Figure – 3 :** Stages of the Mechanical System $[M \mid m_1 \mid m_2]$ throughout experiment.

Defining the state S of the system by the masses present at a particular time, the initial state can be represented by:

$$S_o \equiv \frac{M + m_1 + m_2}{\Delta t} \quad (4)$$

Relation ( 4 ) reproduces the fact that at time Δt all the bodies are together and just come in to observation ( Fig. 3); following this, the stages of the mechanical system [ M, $m_1$ , $m_2$ ] can be represented by similar relations:

$$S_1 \equiv \frac{M + m_2}{\Delta t + \delta t} + \frac{m_1}{-\delta t_1} \quad S_2 \equiv \frac{M + m_1 + m_2}{\Delta t + 2\delta t} \quad S_3 \equiv \frac{M + m_1}{\Delta t + 3\delta t} + \frac{m_2}{-\delta t_2} \quad S_4 \equiv \frac{M + m_1 + m_2}{\Delta t + 4\delta t} \quad (4.1)$$

At the initial stage $S_o$ all the bodies are together, but at the first experimentation (stage $S_1$ ) the body $m_1$ is retired from contact with M and $m_2$ . Then $m_1$ follows its own way and lasts a "proper time" – $\delta t_1$ to reach equilibrium. An analogous situation appears in stage 3 for the body $m_2$. The negative value indicates that these bodies are relaxing back to Δt , or to equilibrium. However, due to the evolution of the system, the falling bodies reach equilibrium at a later time, Δt + nδt (n = 1,2,...), where Δt + nδt is the proper time of the simultaneous bodies and δt the time interval between observations.

Note that δt ≥ $\delta t_{1,2}$ is an experimental condition imposed by the nature of the measurement process: the evolution of the mechanical system must be lower than the "rate of fall" of the (internal) bodies $m_1$ and $m_2$; this represents an uncertainty on the observation of δt carried out by the (internal) observer that measures $\delta t_{1,2}$ .

The change between adjacent steps can be written as :

$$\Delta S_{i,i+1} \equiv S_{i+1} - S_i \quad (5.1)$$

Conversely, figure 3 shows a dynamical evolution of the system [M,$m_1$,$m_2$] whereas each stage represents a static picture of M, $m_1$ and $m_2$; thus, as a result of evolution, each stage can also be represented as:

$$S_i^* \equiv S_i + \Delta S_{i-1,i} \quad (5.2)$$

Except for the initial state $S_o$ , $S_i^*$ can be computed for all stages ( i>0) and represents an excited or meta-stable state resulting from the act of experimentation, as opposed to the static picture. Now, two ways of evolution can be distinguished: evolution 'A', when the first experiment is the "free fall" of body $m_1$ and the second the " free fall " of body $m_2$ and evolution 'B' when the first experiment is the "free fall" of body $m_2$ and the second the " free fall " of body $m_1$ . The changes for these two cases can now be computed.

**1)- Evolution 'A' :**

This evolution follows the picture of Figure 3; computing the changes, the first one gives:

$$\Delta S_{01} \cong \frac{M + m_2}{\Delta t + \delta t} - \frac{m_1}{\delta t_1} - \frac{M + m_1 + m_2}{\Delta t}$$

or,

$$\Delta S_{01} \cong (M + m_2)\left[\frac{1}{\Delta t + \delta t} - \frac{1}{\Delta t}\right] - m_1\left(\frac{1}{\Delta t} + \frac{1}{\delta t_1}\right)$$

The first stage after experimentation becomes:

$$S_1^* \cong (M + m_2)\left[\frac{2}{\Delta t + \delta t} - \frac{1}{\Delta t}\right] - m_1\left(\frac{1}{\Delta t} + \frac{2}{\delta t_1}\right)$$

Similarly, stages 2 and 3 after experimentation are :

$$S_2^* \cong (M + m_2)\left[\frac{1}{\Delta t} + \frac{2}{\Delta t + 2\delta t} - \frac{2}{\Delta t + \delta t}\right] + m_1\left(\frac{1}{\Delta t} + \frac{2}{\Delta t + 2\delta t} + \frac{2}{\delta t_1}\right)$$

$$S_3^* \cong -M\left(\frac{1}{\Delta t} - \frac{2}{\Delta t + \delta t} + \frac{2}{\Delta t + 2\delta t} - \frac{2}{\Delta t + 3\delta t}\right) - m_1\left(\frac{1}{\Delta t} + \frac{2}{\Delta t + 2\delta t} - \frac{2}{\Delta t + 3\delta t} + \frac{2}{\delta t_1}\right) - m_2\left(\frac{1}{\Delta t} - \frac{2}{\Delta t + \delta t} + \frac{2}{\Delta t + 2\delta t} - \frac{2}{\delta t_2}\right)$$

and the final stage of the evolution 'A' pictured in Figure 3 is:

$$S_F^A \equiv S_4^* \cong (M + m_1 + m_2)\left\{\frac{1}{\Delta t} + \frac{2}{\Delta t + 2\delta t} + \frac{2}{\Delta t + 4\delta t}\right\} + 2\frac{m_1}{\delta t_1} + 2\frac{m_2}{\delta t_2} - (M + m_2)\frac{2}{\Delta t + \delta t} - (M + m_1)\frac{2}{\Delta t + 3\delta t} \quad \textbf{(6.A)}$$

**2)- Evolution 'B' :**

Alternatively, the order of experiments can be reverted by exchanging *m₁* and *m₂* ; states $S_o$, $S_2$ and $S_4$ are unchanged, but stages 1 and 3 become:

$$S_1 \equiv \frac{M + m_1}{\Delta t + \delta t} + \frac{m_2}{-\delta t_2} \quad ; \quad S_3 \equiv \frac{M + m_2}{\Delta t + 3\delta t} + \frac{m_1}{-\delta t_1} \quad ;$$

This leads to a final state as:

$$S_F^B \cong (M + m_1 + m_2)\left\{\frac{1}{\Delta t} + \frac{2}{\Delta t + 2\delta t} + \frac{2}{\Delta t + 4\delta t}\right\} + 2\frac{m_1}{\delta t_1} + 2\frac{m_2}{\delta t_2} - (M + m_1)\frac{2}{\Delta t + \delta t} - (M + m_2)\frac{2}{\Delta t + 3\delta t} \quad \textbf{(6.B)}$$

This shows that, depending on the order of experimentation, the final state of the system differs by a quantity:

$$\Delta S_{process} \equiv \left|S_F^A - S_F^B\right| \cong 2 \times \left|(m_1 - m_2)\left[\frac{1}{\Delta t + \delta t} - \frac{1}{\Delta t + 3\delta t}\right]\right| \quad (7)$$

This result is independent of the choice of the light bodies, since no matter $m_1 = m_2 + \delta M$ or $m_2 = m_1 + \delta M$, the value of $\Delta S$ remains the same.

Thus experimentation gives rise to distinct results, according to the order of the procedure. Conversely, the act of experimentation disturbs the system, in a sense that it does not return to the same configuration; the changes related to each of the masses M, *m₁* and *m₂* can also be calculated. Choosing evolution 'B' , the last (accumulated) change for each body gives:

**1)- Body - M**

$$\Delta S_M \cong M\left(\frac{1}{\Delta t} - \frac{2}{\Delta t + \delta t} + \frac{2}{\Delta t + 2\delta t} - \frac{2}{\Delta t + 3\delta t} + \frac{1}{\Delta t + 4\delta t}\right)$$

Now this result can be written as two sets,

$$\Delta S_M \cong M\left(\frac{1}{\Delta t} - \frac{1}{\Delta t + \delta t} + \frac{1}{\Delta t + 2\delta t} - \frac{1}{\Delta t + 3\delta t} + \frac{1}{\Delta t + 4\delta t}\right) - M\left(\frac{1}{\Delta t + \delta t} - \frac{1}{\Delta t + 2\delta t} + \frac{1}{\Delta t + 3\delta t}\right)$$

each one – according to Leibniz's [2] theorem for alternating sets – gives a positive sum not superior to the first term, that is :

$$0 \leq \Delta S_M \leq M\left(\frac{1}{\Delta t} - \frac{1}{\Delta t + \delta t}\right)$$

Thus a rough result is:

$$\Delta S_M \cong M\left(\frac{1}{\Delta t} - \frac{1}{\Delta t + \delta t}\right) \quad (8.1)$$

**2)- Bodies - $m_1$ and $m_2$**

$$\Delta S_{m_1} \cong m_1\left(\frac{1}{\Delta t} + \frac{2}{\Delta t + 2\delta t} + \frac{1}{\Delta t + 4\delta t} + \frac{2}{\delta t_1} - \frac{2}{\Delta t + 3\delta t}\right)$$

as above,

$$\Delta S_{m_1} \cong m_1\left(\frac{1}{\delta t_1} + \frac{1}{\Delta t} + \frac{1}{\Delta t + 2\delta t} - \frac{1}{\Delta t + 3\delta t}\right) + m_1\left(\frac{1}{\delta t_1} + \frac{1}{\Delta t + 2\delta t} - \frac{1}{\Delta t + 3\delta t} + \frac{1}{\Delta t + 4\delta t}\right)$$

Thus a rough result is:

$$0 \leq \Delta S_{m1} \leq \frac{2m_1}{\delta t_1}$$

Here the series theorem argument is irrelevant once the first term dominates the set due to the smallness of $\delta t_1$.
Similarly, $m_2$ gives :

$$0 \leq \Delta S_{m2} \leq \frac{2m_2}{\delta t_2}$$

The right hand of these last relations give:

$$\frac{\delta t_1}{2} \Delta S_{m1} \leq m_1 \quad (8.2)$$

$$\frac{\delta t_2}{2} \Delta S_{m2} \leq m_2 \quad (8.3)$$

Recovering that the quotient $m_1 / m_2$ was already determined by experiment, furnishing the usual ratio of accelerations (1), the last two relations give:

$$\frac{m_1}{m_2} \cong h\frac{\dot{v}_2}{\dot{v}_1} \geq \frac{\Delta S_{m_1} \delta t_1}{\Delta S_{m_2} \delta t_2} \quad (9)$$

where $h$ is the 'constant of proportionality' relating the ratio of masses to the ratio of accelerations – equation (3).
Thus it is possible to establish a connection between the motion of the masses and the resulting order of the parts:

$$\frac{\Delta S_{m_1}}{\Delta S_{m_2}} \approx h(\alpha) \frac{\langle v_2 \rangle}{\langle v_1 \rangle} \quad (10)$$

where $\langle v_i \rangle \equiv \delta t_i \, \dot{v}_i$ is the measured (mean) velocity of $m_i$, since $\delta t_i$ is the experimental time of fall associated to body $m_i$.

# 4 – *Comments*

Equation (10) has a form similar to equation (1): velocities take the place of accelerations and the ratio of masses are substituted by the 'ratio of states'; also, the motion of the [$m_1$ | $m_2$] subsystem probably occurs in the direction of minimum relative $\Delta S$ changes and equations (8), (9) and (10) seems to conform with the Principle of Inertia. Moreover, though equations (8.2) and (8.3) were written in terms of the variables of the bodies $m_1$ and $m_2$, equation (10) inserts the *h* function as a connection with the motion of the system as a whole; this opens the possibility to interpret the different values of *h* as distinct paths that the light bodies ($m_1$, $m_2$) may follow on their way back to equilibrium, a 'classical' uncertainty in the motion of the mechanical system. Conversely, this function is directly related to the picture of Newtonian mechanics represented by figure 3, so the explicit introduction of the time of the system in the Classical formalism requires an analysis of the meaning of δt in the context of indetermination.

Nature registers the existence of three distinct temporalities and strongly indicates a fourth component, while physics deals with just two. The temporalities of physics are the system's time and sensor's time; that of Nature are past, present and future; the fourth component is *atemporality*.

*Atemporality* does not belong to the nature of measurable effects, but to the reality of facts; also, it does not mean that there is no time but that all times collapse. Thus, there is a primordial indeterminacy on describing the facts of reality by means of the effects of nature. Moreover, physical systems are taken as evolving in time (external parameter) but detected on time, or at time, by a detector somehow connected to the system (internal parameter).

The condition $\delta t \geq \delta t_{1,2}$ reflects an uncertainty between facts and effects, observation of the motion of the system and measurement or determination of the motion of a part of the system, $m_1$ or $m_2$. $\delta t < \delta t_{1,2}$ implies an indeterminacy on the measurement of the motion of the bodies $m_1$ and $m_2$; $\delta t = 0$ implies no observation of the motion of the system [ M, $m_1$, $m_2$ ], since both the variation of the process (7) and that of the heavier body (M) (8.1), present a null result.

In fact, the evolution of the system [ M, $m_1$, $m_2$ ] is an attribute of external observation and can only be sensed by an internal observer by means of the internal fluctuations of the system; thus, the time δt can only be estimated by the measurement of $\delta t_{1,2}$. In this sense, $\delta t_{1,2}$ is a *measurable quantity* while δt is an *observable entity*. That is to say, the *experimental* point of disagreement among the branches of mechanics, setting back a division between experimental entities – the ones resulting from detectable *effects of nature* – and 'just' observable entities, those associated to *facts of reality*, whose detection demands a conceptual operation.

The question of *observability* [3] plays a central hole in mechanics and generates many debates both at the classical end but mainly at the quantum and relativistic levels; in particular, observability of time is questioned so that quantities that change with time are also under suspicion. On the other hand, it is argued that observable is a (relative) quantity expressing a balance (correlation) among dynamical variables of the system. The overall discussion has to do with localisation in space and also in time; but localisation in time seems unrealistic once time is *in side* physical bodies; perhaps on time, or at time. This way, sequencing of events – the practical version of *Causality* – appears as the conceptual (erroneous/erratic) counterpart of time observability, since it involves time uncertainty and/or time dilation.

Note, also, that equation (10) correlates observable entities – the ratio of states – to measurable quantities, the ratio of velocities. Further, the conceptual notion of Causality becomes erratic, once $h$, apart from the ambiguity, is a function of the (internal) dynamical variables of the system.

The classical formulation of mechanics does not require the concept of *proper time* $\delta t_{1,2}$ neither that of the external time $\delta t$, explicitly. In such a context, neglecting the changes of the system, observability is set inside the system and mixed with experimentation so that the results reveal a linear relationship between the bodies subjected to the experiment. From the perspective of an external observer – the one that pictured (photographed) the stages of Fig. 3 – observation and measurement are quite distinct once they are not exactly at the same time, that is, in the same place.

Relativistic mechanics provides a solution for this problem by transporting the data of the experiment to the observer, at the cost of a time dilation; quantum mechanics solves the problem at the cost of momentum–position / time-energy uncertainty. Here the problem was settled by removing the constraint $\Delta t = \delta t_{1,2} = \delta t$. The implicit argument is that the $\delta t_{1,2} = \delta t$ constraint can be removed by quantum mechanical methods while the $\Delta t = \delta t_{1,2}$ by relativistic ones.

The correlation among $\delta t$, $\Delta t$ and $\delta t_{1,2}$ appears as a function $h$, which is a relation among dynamical variables of the system; more precisely, the quantity **S** originally defined in terms of proper times. Conversely, definition of time is state dependent once there is an uncertainty on determination of states; thus, time is determined only statistically (thermodynamically or statistical mechanics sense) and the only temporality that rests at the mechanical level is atemporality; the proper time in physical bodies appears only connected to the motion ( action ).

Solid state physics deals with the notion of carriers, charge-carriers, such as electrons and holes in the theory of semiconductors; analogously, mechanics can say that physical bodies are time–carriers. Removing body $m_1$ from contact with bodies M and $m_2$ means subtraction of a quantity $\delta t_1$; return of the time-carrier $m_1$ to contact with the time reservoir ($\Delta t + n\delta t$) represented by M (and $m_2$) means that a time $\delta t_1$ has elapsed and that the $m_1$ body lost his proper time.

Finally, the quantity **S** defined by (4) represents – according to (4.1) – the evolution of the system. Moreover, the change in **S** between any two equilibrium states (equations 5.1,2 / 6.A,B) is found by taking the system along the path connecting the states, dividing the mass added to the system at each point of the path by the time of the system and summing the quotients thus obtained. Thus a quantity given by

$$\sum \frac{\delta M}{t}$$

is equal to the difference between the values of the function **S** at the end points of the path. This function has all the ingredients attributed to Entropy. Thus, assuming that the definition stated by (4) do represent the state of the mechanical system [ M, $m_1$, $m_2$ ], the quantity **S** should be considered as a 'classical' Entropy, neglected in the original Newtonian formalism of mechanics.

Also, considering the reality of the motion (V) of bodies (M) throughout space (ΔX), the quantity **S** can be written as MV/ΔX, that is:

$$S = \dot{M} = \frac{M}{\Delta X / V}$$

Thus, tough the abstract entity named time appears useful for measurable purposes and physical descriptions such as equation (4) or figure 3, explicit computations show that the real physical flow can be settled in terms of the dynamical variables of the system.

------ *** ------

# 5 – *Conclusions*

This contribution reviews the original experiment (formulation) of Classical Mechanics centring the discussion on Indeterminacy instead of Causality. Analysis of a more realistic experiment ( Fig. 3 ) leads to multiple solutions, expressed by the values of *h(α)*, revealing that the classical results are based on a *gedanken* experiment and on a single data ($\alpha \rightarrow 0$). Application of a simple description of the motion, without any attempt to quantification other than the original Newtonian results, indicates that the act of experimentation disturbs the evolution of the system.

Employing the concept of Mechanical States, those distinct results can be associated to variations of the order (Entropy) of the mechanical system [ M| $m_1$| $m_2$]. This demands for an explicit distinction between facts and effects, or removal of the constraint between observable entities and measurable quantities.

Expressed in terms of time – proper times and change of rate of the system – the constraint was removed employing quantum and relativistic arguments based on a statistical (thermodynamically sense) notion of time. Finally, since any representation of physical reality must be settled in terms of the dynamical variables of the system, the possibility to write the quantity **S** in terms of masses, velocities and space was shown.

Development of a formalism centred on the arguments here described, mainly the concept of *State/Entropy*, may open the possibility to an uniform synthesis of the branches of mechanics, provided that a comprehensive method of motion detection can be envisaged.

------ *** ------

# 6 – *References*

# 7 – *Acknowledgements*

Discussions with A J Roberto Jr., L F Delboni, M M H Barreira and W. Amstalden are gratefully acknowledged. Comments/suggestions are welcomed and can also be sent to assump@fem.unicamp.br.